\documentclass[]{spie}  

\usepackage{amsmath,amsfonts,amssymb}
\usepackage{graphicx}
\usepackage[colorlinks=true, allcolors=blue]{hyperref}

\newcommand{\arcsec}{''}

\title{Laboratory testing and characterization of a hybrid fast/slow readout mode for the H2RG detectors in SCALES}

\author[a]{Peyton Benac}
\author[a]{Michael P. Fitzgerald}
\author[a]{Eric Wang}
\author[a]{Kenneth Magnone}
\author[a]{Chris Johnson}
\author[b]{Andrew J. Skemer}
\author[c]{Philip Hinz}
\author[c]{Nick MacDonald}
\author[c]{Dale Sandford}
\author[c]{R. Deno Stelter}
\author[c]{William Deich}
\author[d]{Brittany E. Miles}
\author[b]{Steph Sallum}
\author[e]{Thomas Greene}
\author[f]{Markus Loose}
\author[g]{Richard Blank}

\affil[a]{University of California Los Angeles, 475 Portola Plaza, Los Angeles, California, USA}
\affil[b]{University of California Santa Cruz, 1156 High Street, Santa Cruz, California, USA}
\affil[c]{University of California Observatories, 550 Red Hill Rd, Santa Cruz, California, USA}
\affil[d]{University of Arizona, 933 N Cherry Ave, Tucson, Arizona, USA}
\affil[e]{NASA Ames Research Center, Moffett Blvd, Mountain View, California, USA}
\affil[f]{Markury Scientific Inc, Thousand Oaks, California, USA}
\affil[g]{AstroBlank Scientific LLC, Ventura, California, USA}

\authorinfo{Further author information: (Send correspondence to P.B.)\\P.B.: E-mail: pbenac@g.ucla.edu }

\begin{document} 
\maketitle

\begin{abstract}

The upcoming SCALES (Slicer Combined with Array of Lenslets for Exoplanet Spectroscopy) instrument for W.M. Keck Observatory will enable new imaging and low-/mid-resolution coronagraphic integral field spectroscopic observations over the wavelength range from 2-5 microns. At the heart of the instrument are two HgCdTe Teledyne Imaging H2RG detectors, designed for a 100kHz pixel clock rate (slow mode) with a fixed 4-channel readout. However, in ground-based operation at these wavelengths, the resulting minimum frame readout time will result in the infrared background saturating the detector. To enable high quality observations without saturation from the bright infrared sky background, we operate the detectors using a custom cable for buffered readout via the Teledyne Imaging SIDECAR ASIC followed by an AstroBlank/Markury Scientific MACIE controller card controlled by custom firmware. This combination allows the detector to be read out at faster pixel clock rates. This, in combination with the slow-mode H2RG, is what we characterize as hybrid fast-slow readout, enabling readout up to 18 times faster than would be possible in slow mode alone. In the UCLA Infrared Lab, we have performed room-temperature and cold tests with the H2RG detectors. We test and optimize full-frame data acquisition with pixel clock rates from 0.2-1.8 MHz.
In these proceedings, we present a summary of the controller software used to operate the H2RG-ASIC-MACIE system. We present the methodology of, and preliminary results from, the UCLA tests of cryogenic operation of both H2RG detectors. We also outline the next steps in verification of detector performance, as well as integration with the SCALES instrument.

\end{abstract}


\section{INTRODUCTION}
\label{sec:intro}  
Current ground-based searches to directly detect and characterize exoplanets have difficulty detecting cooler planets, limiting our ability to study the later phases of planet formation and planets located further from their host stars. Extending methods that are well established in the near infrared into longer wavelengths will enable us to extend direct planet detection and characterization to cooler temperature regimes. This is one of the chief motivations for the creation of a new instrument for the Keck II telescope at Keck Observatory, SCALES. 
SCALES (Slicer Combined with Array of Lenslets for Exoplanet Spectroscopy) instrument is a coronagraphic integral field spectrograph with an additional imaging arm.
The SCALES spectrograph operates at wavelengths from 2 to 5 microns, enabling unique science cases and serving as a complement to high-contrast instruments assisted by extreme adaptive optics (AO) systems like GPI, SPHERE, and CHARIS, which all operate from 1 to 2.5 microns. It is also complementary to  existing Keck instruments OSIRIS and NIRSPEC, neither of which provides IFS observations at the SCALES wavelength range, or has coronagraphs optimized for exoplanet observations. The SCALES imaging arm, which operates from 1 to 5 microns, will also serve as a replacement in part for the current NIRC2 imager. SCALES uses a combination of a slicer and a lenslet array to more efficiently utilize the detector pixels, compared to lenslet-only IFS like OSIRIS \cite{Larkin_SPIE_2006, Stelter_SPIE_2022}. SCALES has three main observational modes: low-resolution integral field spectroscopy (IFS), with spectral resolution from 35 to 200, medium-resolution IFS with spectral resolution from 3,000 to 7,000, and imaging a 12\arcsec\  by 12\arcsec\ field of view with a sampling of 6 mas \cite{skemer_spie_22}. 
The SCALES instrument development is led out of UC Santa Cruz by principal investigator Andy Skemer, instrument scientist Deno Stelter, and project scientist Steph Sallum. The development of SCALES involves the contribution of many teams working on individual subsystems, which are being integrated at UCSC. 

Infrared observations from ground-based astronomical observatories face distinct challenges, including greatly increased sky brightness compared to visible light observations. For the specific case of SCALES, this problem is exacerbated by unique properties of the HgCdTe Teledyne Imaging H2RG detectors provided by NASA for use in this project. These detectors were manufactured in the batch of detectors produced for JWST/NIRSPEC, and were therefore optimized for the readout architecture used in JWST. Specifically, these detectors were hard-wired to use 4 output channels and the slow-mode analog preamplifiers, with a pixel clock rate of about 100kHz, which corresponds to a minimum frame time of approximately 10.5 seconds. For a ground-based instrument, we need to be able to increase the detector clocking rate to decrease the minimum frame time, allowing us to observe brighter sources without saturating, to minimize the impact of the sky brightness on our science results. 
To enable this, the SCALES team works with Markury Scientific Inc and AstroBlank LLC to utilize the MACIE controller card with customized firmware to operate our H2RG detectors at faster clock rates, accomplished by using the fast-mode analog-to-digital converters within the SIDECAR ASIC. Additionally, to operate in this mode, we also use custom cables between the H2RG and ASIC, which include resistors to allow buffered circuit readouts that are necessary for this hybrid operational mode. This allows us to operate the detector at pixel clock rates greater than 1 MHz, resulting in minimum frame times of 1.05 seconds or less. 
\begin{figure}
    \centering
    \includegraphics[width=\linewidth]{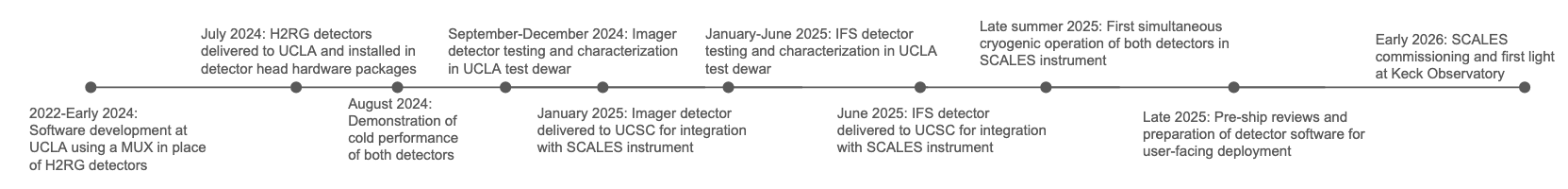}
    \caption{A timeline summarizing major recent milestones in the detector package preparation for SCALES.}
    \label{fig:Timeline}
\end{figure}

In the UCLA Infrared Lab, we worked to test this custom firmware and develop the detector controller software. To accomplish this, we began by testing with a MUX in place of the real H2RG detector. Figure \ref{fig:Timeline} shows the timeline for work on the SCALES detector heads from 2022 to present; this software development work was the primary focus until summer 2024, when the H2RG detectors were delivered to UCLA and integrated into their detector head housing hardware for further testing. The detailed mechanical design of the detector heads is generally beyond the scope of this work (some detail is provided in Section \ref{sec:hardware} and figures therein), but it is worth noting that the two systems are designed to minimize the possibility of thermal or electrical shorts that could affect detector performance, and we have demonstrated satisfactory thermal control of the detectors, ASICs, and other hardware during our tests in the UCLA test cryostat. The detector head assemblies can be seen in Figures \ref{fig:ImagerCAD}, \ref{fig:imagerPic},  \ref{fig:IFSPic}, \ref{fig:transpIFS}, and \ref{fig:CADIFS}, within Section \ref{sec:hardware}.

The format of this paper is as follows: Section \ref{sec:hardware} provides a brief description of the SCALES detector hardware. Section \ref{sec:software}  describes the software architecture of the controller code written to interface with the MACIE API, and the basic procedure for operating this software. Section \ref{sec:tests} describes the test procedures used to characterize both H2RG detectors in the UCLA cryostat. Section \ref{sec:results} presents the visual and statistical characterization of the detector performance during these tests, as well as a brief description of the analysis tools used for this characterization. Section \ref{sec:disc} connects this characterization to the next steps for the SCALES instrument.

\section{DETECTOR HARDWARE}
\label{sec:hardware}

\begin{figure}
    \centering
    \includegraphics[width=\linewidth]{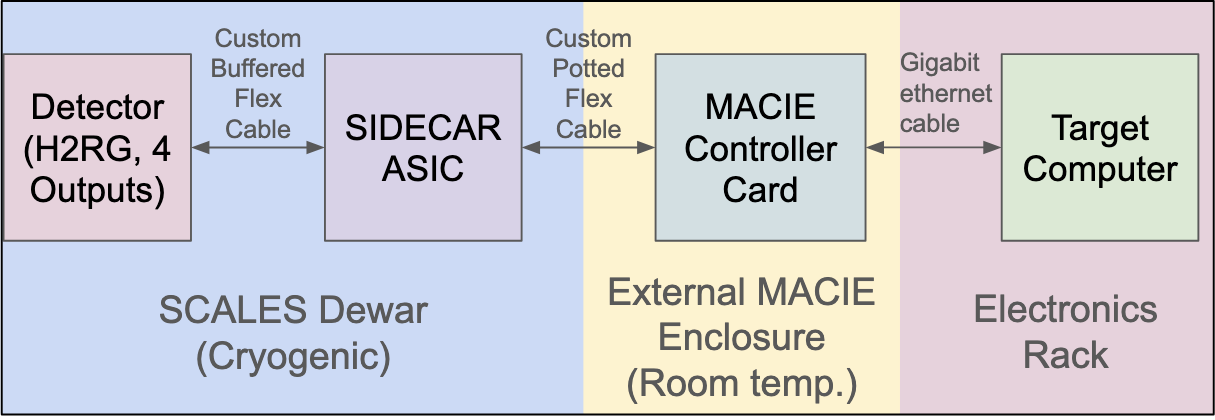}
    \caption{A block diagram summarizing the basic relationships between the electronic hardware for each of the 2 SCALES detectors.}
    \label{fig:hw}
\end{figure}

Each H2RG detector is connected via a custom buffered flexible cable within its detector head to a SIDECAR ASIC, which is then connected with a custom potted flexible cable to the MACIE card that is operated at room temperature. A block diagram of this setup is presented in Figure \ref{fig:hw}. 
\subsection{H2RG Detectors}
The Teledyne HAWAII-2RG infrared arrays were provided by NASA to SCALES to support its science mission. These HgCdTe (MCT) arrays feature 2048x2048 pixels (2040x2040 photosensitive, with 4 reference pixels on each external edge) at an 18 $\mu$m pitch \cite{Blank_2011}. These pixels are read out via 4 separate outputs. We have two separate detectors, one for the imaging arm of SCALES and one for its spectrograph (IFS). 
Teledyne HxRG detectors can run in slow mode, with a typical speed of 100 kilopixels per second per output, or in fast mode, with a speed of 2 megapixels per second per output \cite{skemer_spie_22}. In fast mode operation, a generic H2RG full frame can be read out in as little as 0.065 seconds. However, the JWST flight spare detectors allocated for use in SCALES have been hardwired to use 4 outputs in slow mode, which reduces both the noise and the energy consumed. However, the use of only 4 readouts (as opposed to 32 readouts, which is also possible with HxRG detectors) reduces the maximum readout speed by a factor of 8. to avoid saturating with the bright infrared sky background, we must achieve faster readout. This is accomplished by operating in a new hybrid buffered fast-slow mode, where the SIDECAR ASIC analog-to-digital converters operate in fast mode to record frames from the slow-mode-only detector. This allows us to read out full frames sufficiently fast to avoid saturating from the sky background. Additionally, subframes can be read out faster, allowing brighter targets to be observed. To operate in this mode, we use a custom cable that includes extra resistors to allow for buffered readout from the H2RG to the ASIC.

\subsection{SIDECAR ASIC}
The ASIC manages the clocking patterns for detector readout and retrieves data from the detector. The SIDECAR ASICs used in SCALES are focal plane array control electronics produced by Teledyne Imaging Sensors. The SIDECAR ASIC is an ``all in one" CMOS (complementary metal-oxide-semiconductor) micro-chip optimal for use in astronomical applications because it is small in size, highly programmable, and operable down to 30 K \cite{Chen_SPIE_2014}.  The SIDECAR ASIC is used in the Hubble Space Telescope and the James Webb Space Telescope \cite{Dorn_SPIE_2008}.  Within SCALES, the SIDECAR ASIC is located within each of the two detector enclosures (referred to as `detector heads'). Each ASIC is controlled with tuned, custom firmware, and connected to the H2RG with a buffered cable, to enable the optimized hybrid fast-slow operation of the SCALES system.

\subsection{MACIE}
\begin{figure}
    \centering
    \includegraphics[width=0.5\linewidth]{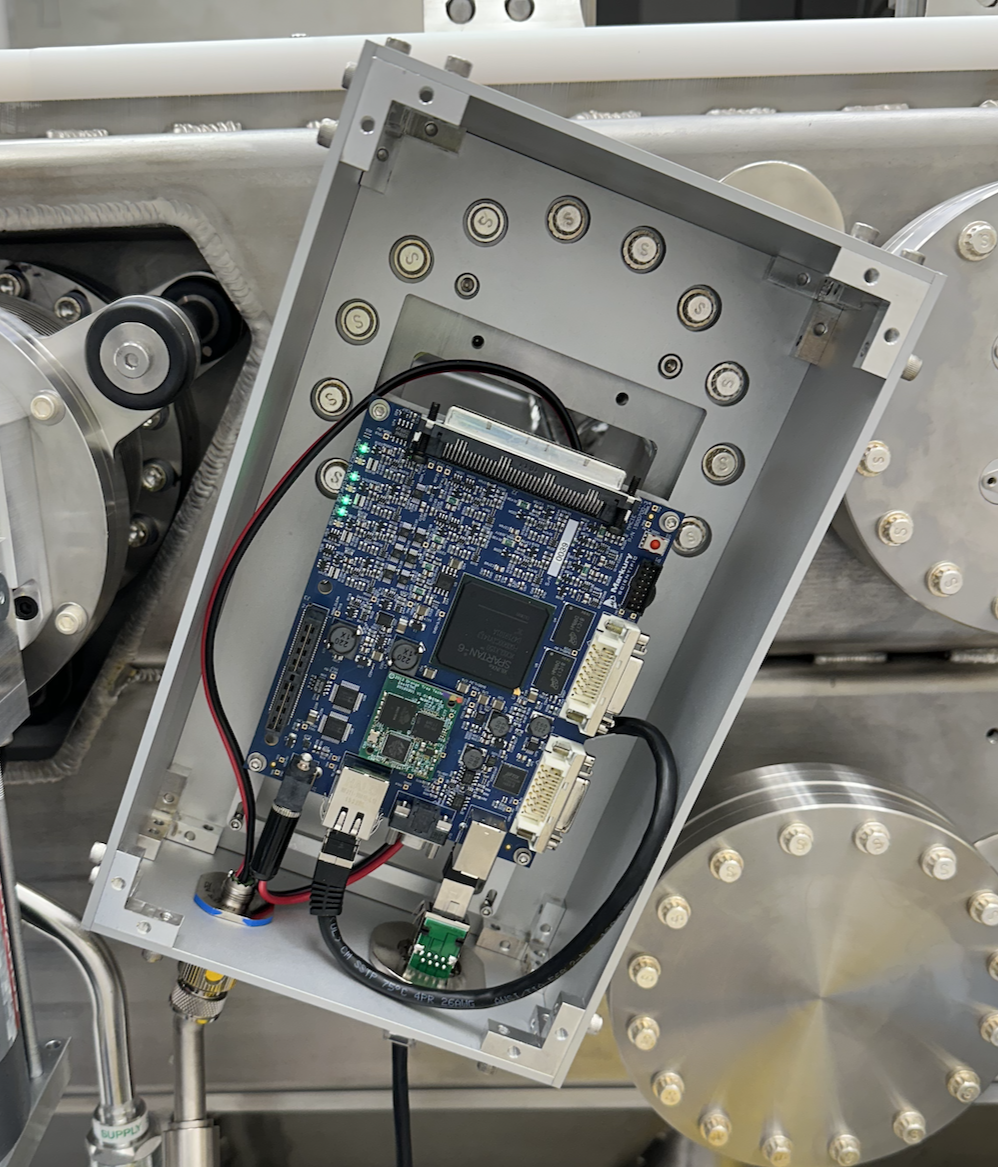}
    \caption{MACIE enclosure for the Imager detector, attached to the exterior of the SCALES cryostat in the UCO lab. The enclosure is shown here without its protective aluminum cover plate. An equivalent enclosure exists on the opposite side of the cryostat, attached to the IFS detector.}
    \label{fig:macieenc}
\end{figure}

The AstroBlank/Markury Scientific MACIE (Multipurpose ASIC Control \& Interface Electronics) controller card is a room-temperature FPGA-based device that serves as the interface between the ASIC/detector and the controlling computer. The MACIE is connected to a 5V power source and a gigabit ethernet connection. Controllable via either a proprietary GUI or via a C/C++ API, the MACIE provides a flexible means of communicating with the detector hardware.  In the completed SCALES instrument, each MACIE card is held in a room temperature enclosure attached to the exterior of the SCALES cryostat, as shown in Figure \ref{fig:macieenc}. 

\subsection{Detector Head Assembly and Validation}
We assembled the detector head hardware at UCLA during July 2024. Assembly procedures were practiced without static-sensitive hardware to minimize risk of electrostatic discharge damaging the sensitive H2RG and ASIC. After the detector heads were assembled, they were checked for thermal or electrical shorts via a multimeter. Electronic readout was validated with room-temperature testing  before beginning any cryogenic tests (see Section \ref{ss:cryo}). 

\subsubsection{Imager Detector Head}
\begin{figure}
    \centering
    \includegraphics[width=0.5\linewidth]{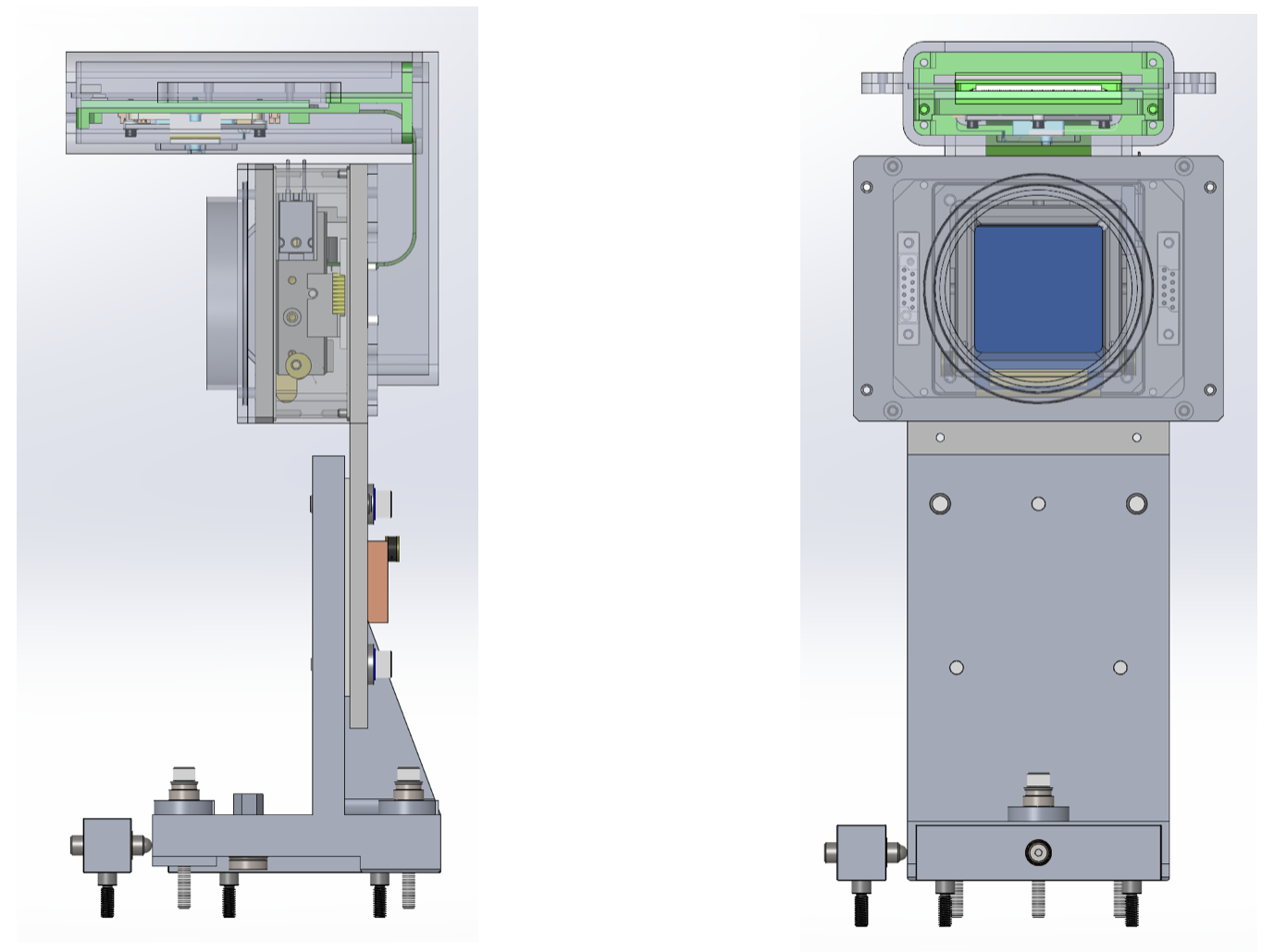}
    \caption{SolidWorks Model of the Imager Detector Head. The detector face is shown in blue in the right panel and the ASIC's insulating G10 frame is highlighted in green.}
    \label{fig:ImagerCAD}
\end{figure}

\begin{figure}
    \centering
    \includegraphics[width=0.4\linewidth]{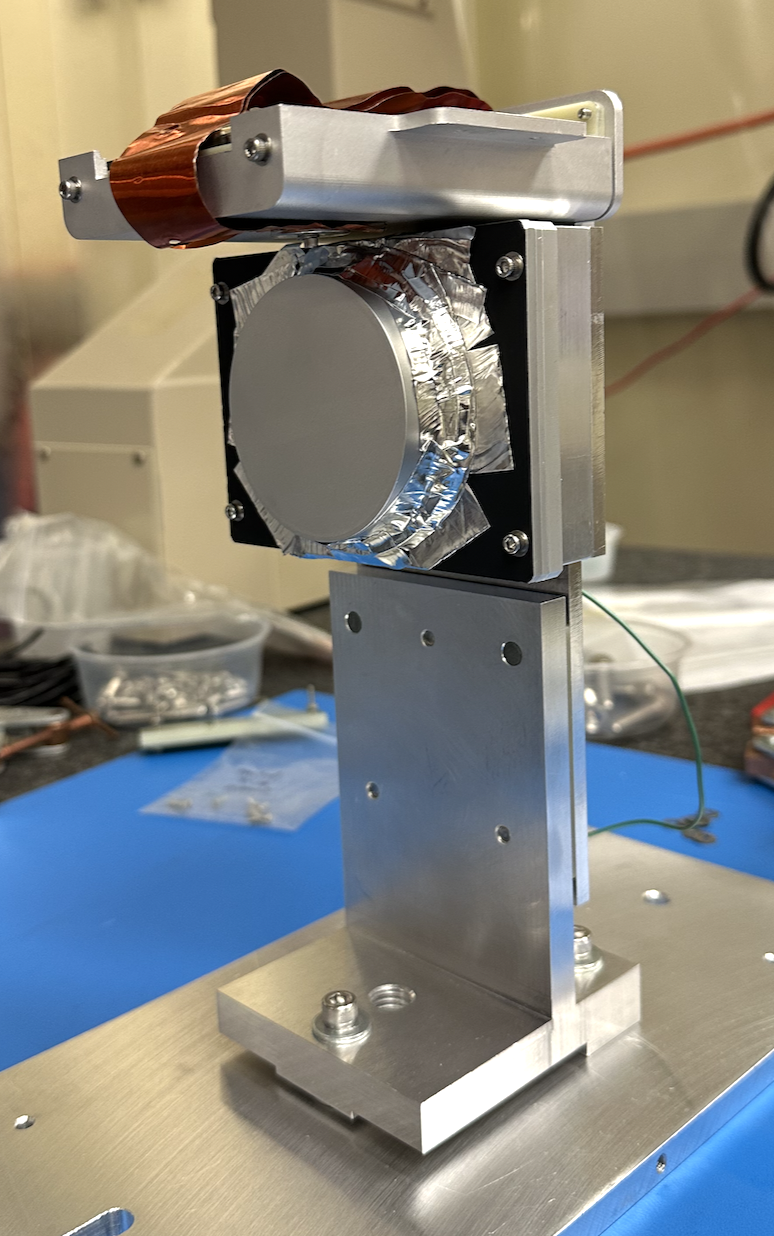}
    \caption{The Imager detector head as assembled in the UCLA IR Lab. The ASIC enclosure (top) is shown without its top cover, but in the completed assembly it is enclosed. The copper attached to the ASIC is for thermal strapping during cryogenic operation.} 
    \label{fig:imagerPic}
\end{figure}

The mechanical design of the Imager detector head features a circular baffle at the detector face and the ASIC enclosure located on top of the detector. This is shown in Figure \ref{fig:imagerPic}. The CAD model of the Imager detector head is also shown in Figure \ref{fig:ImagerCAD}.
 
\subsubsection{IFS Detector Head}
\begin{figure}
    \centering
    \includegraphics[width=0.4\linewidth]{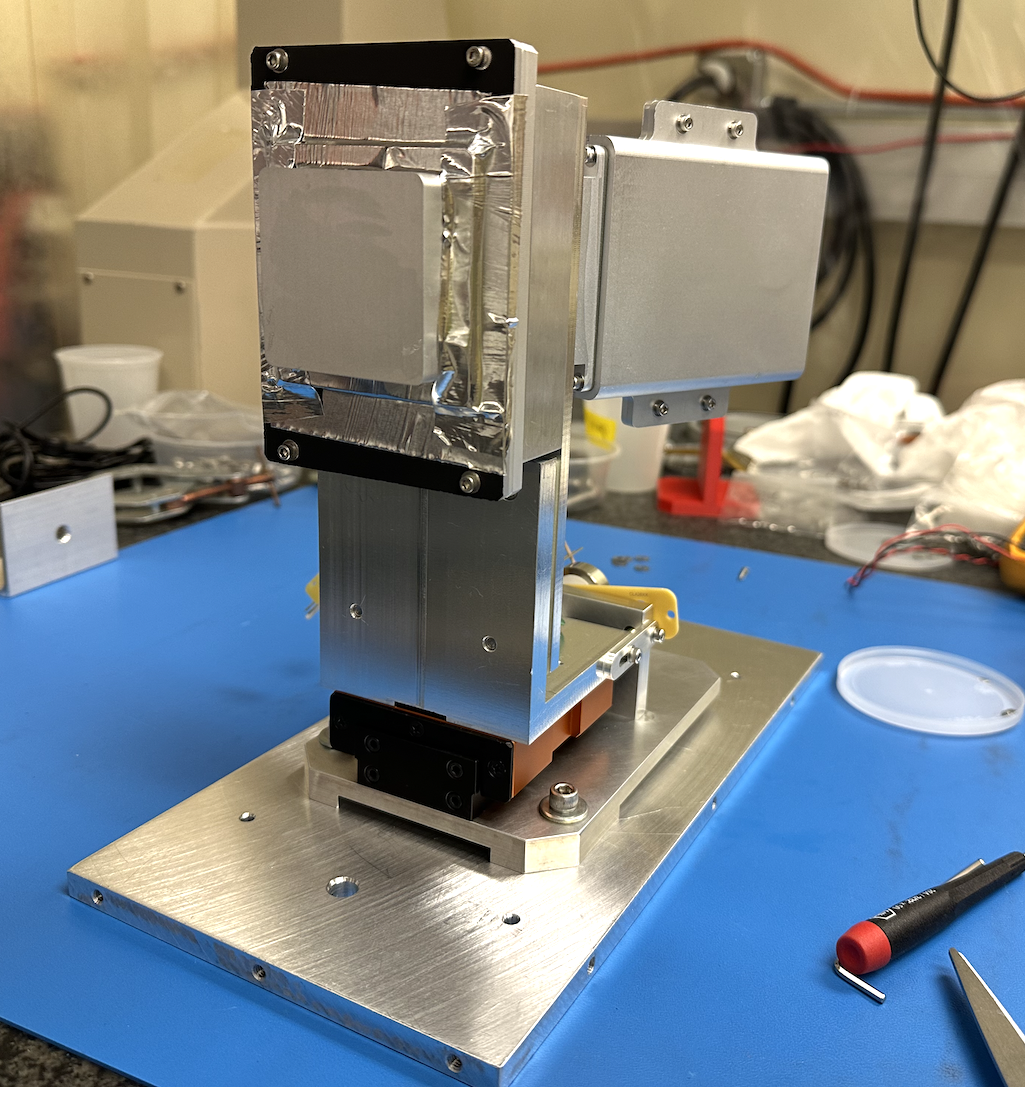}
    \caption{The IFS detector head as assembled in the UCLA IR Lab. The focus stage (black and orange rectangular box) is located under the detector head. The ASIC is located in the aluminum enclosure behind the detector.}
    \label{fig:IFSPic}
\end{figure}

\begin{figure}
    \centering
    \includegraphics[width=0.5\linewidth]{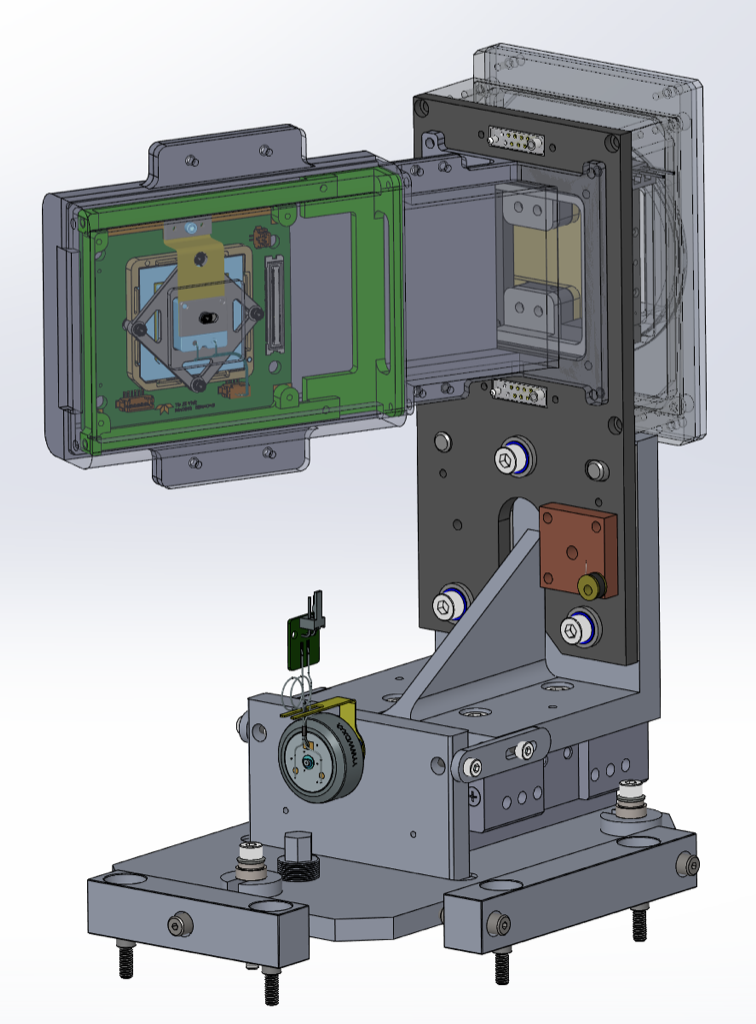}
    \caption{Back view of the SolidWorks Model of the IFS Detector Head. The ASIC is shown in this transparent view as well as its insulating G10 frame (shown in bright green.)}
    \label{fig:transpIFS}
\end{figure}

\begin{figure}
    \centering
    \includegraphics[width=0.5\linewidth]{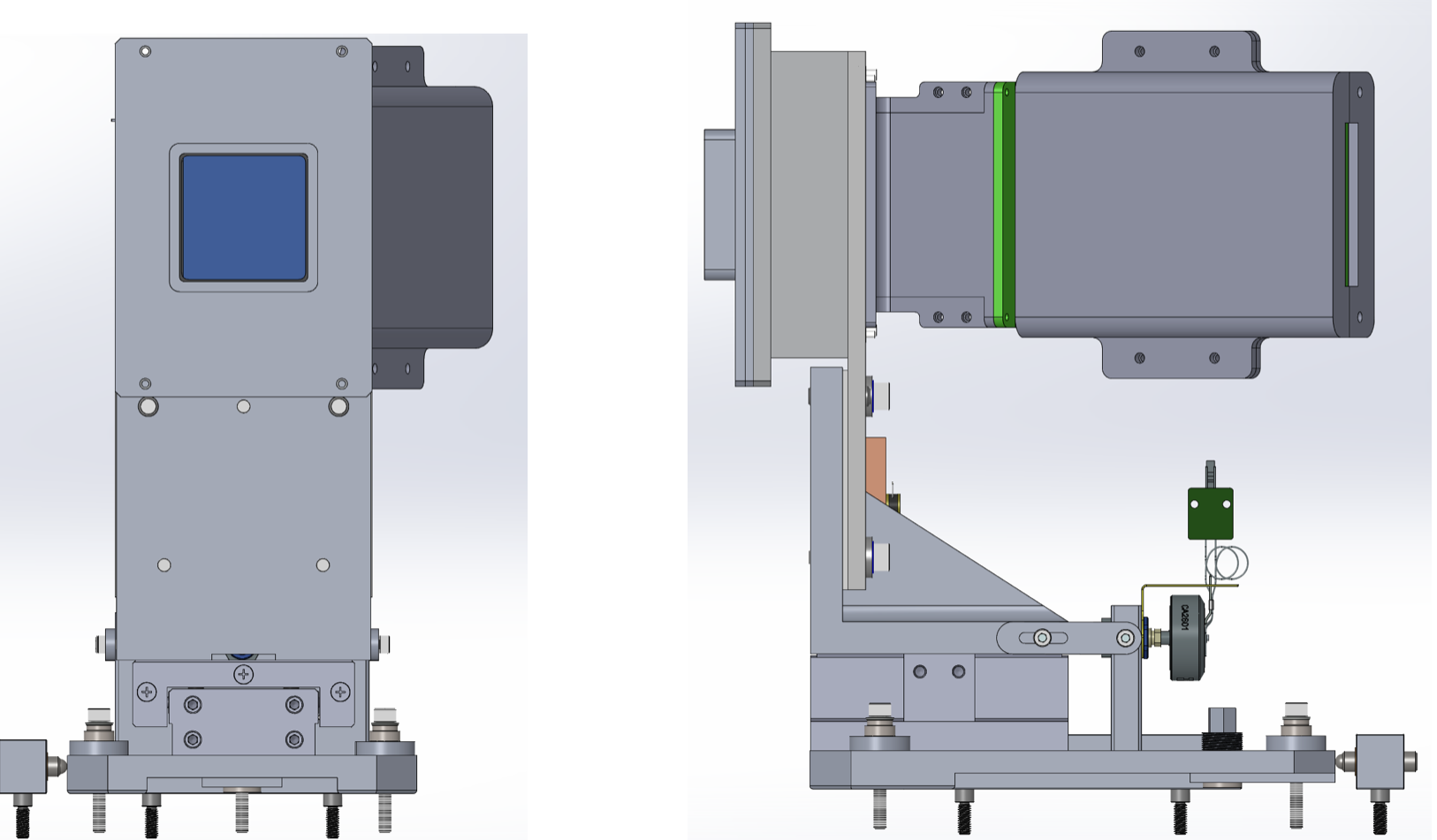}
    \caption{Additional views of the SolidWorks Model of the IFS Detector Head. The photosensitive detector face is shown in blue in the left panel. }
    \label{fig:CADIFS}
\end{figure}

The mechanical design of the IFS detector head features a square baffle at the detector face and the ASIC enclosure extending behind the detector at an angle, as shown in Figure \ref{fig:IFSPic}. CAD models of the IFS detector head are also shown in Figures \ref{fig:transpIFS} and \ref{fig:CADIFS}. To allow for focusing, the IFS detector is mounted atop a linear stage that is driven by a cryoactuator. 

\section{SOFTWARE DEVELOPMENT}
\label{sec:software}
\begin{figure}
    \centering
    \includegraphics[width=\linewidth]{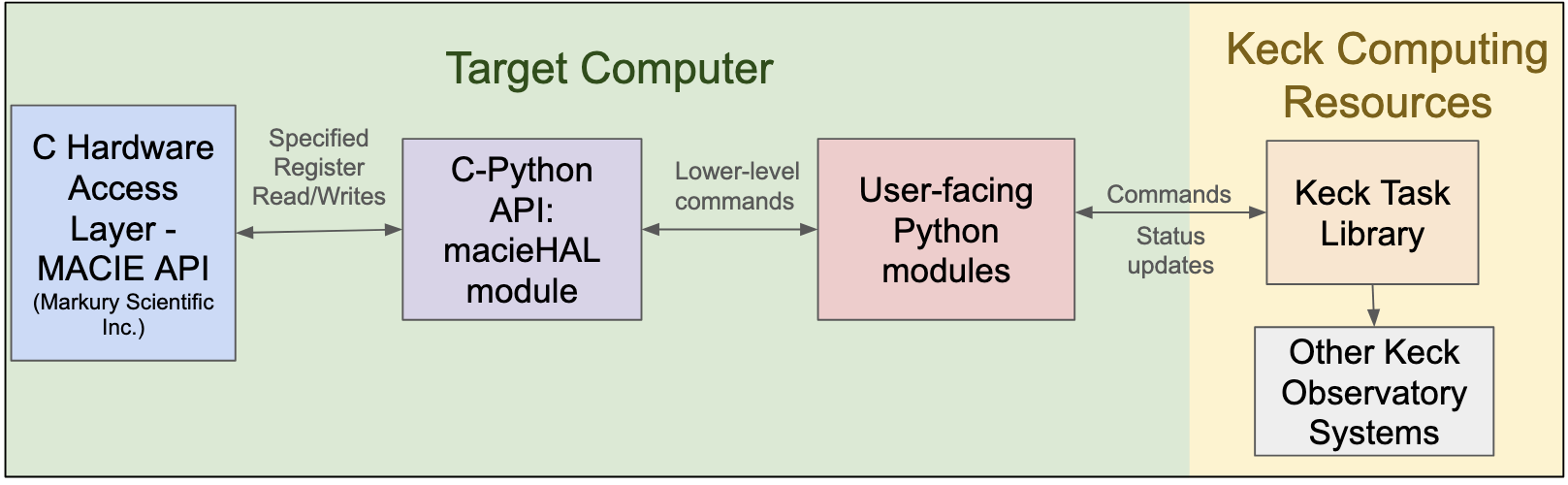}
    \caption{Block diagram outlining the structure of the SCALES detector controller software.}
    \label{fig:software}
\end{figure}

The basic structure of the control software is shown in Figure \ref{fig:software}. Commands (including reading/writing relevant registers for both the SIDECAR ASIC and MACIE, setting acquisition parameters, triggering acquisitions, and writing acquired data to FITS files) are performed via the MACIE C API provided by Markury Scientific and described in detail in their documentation.  An additional C-Python layer was written during the period of 2022-present, with the C library inheriting from example code provided by Markury Scientific. This Python layer will then follow the structure of heritage code in use at Keck Observatory to interface with other observatory systems via the KTL (Keck Task Library) interface \cite{Lupton_ASPC_1993_KTL}. 

Additionally, many tests performed at UCLA were performed or validated against data obtained with the MSAC GUI provided by Markury Scientific. Data and results shown in the following sections will specify whether they were obtained via the Python interface or via the GUI.

\section{TEST PROCEDURES}
\label{sec:tests}
\subsection{MUX Testing} \label{ss:Mux}
\begin{figure}
    \centering
    \includegraphics[width=0.6\linewidth]{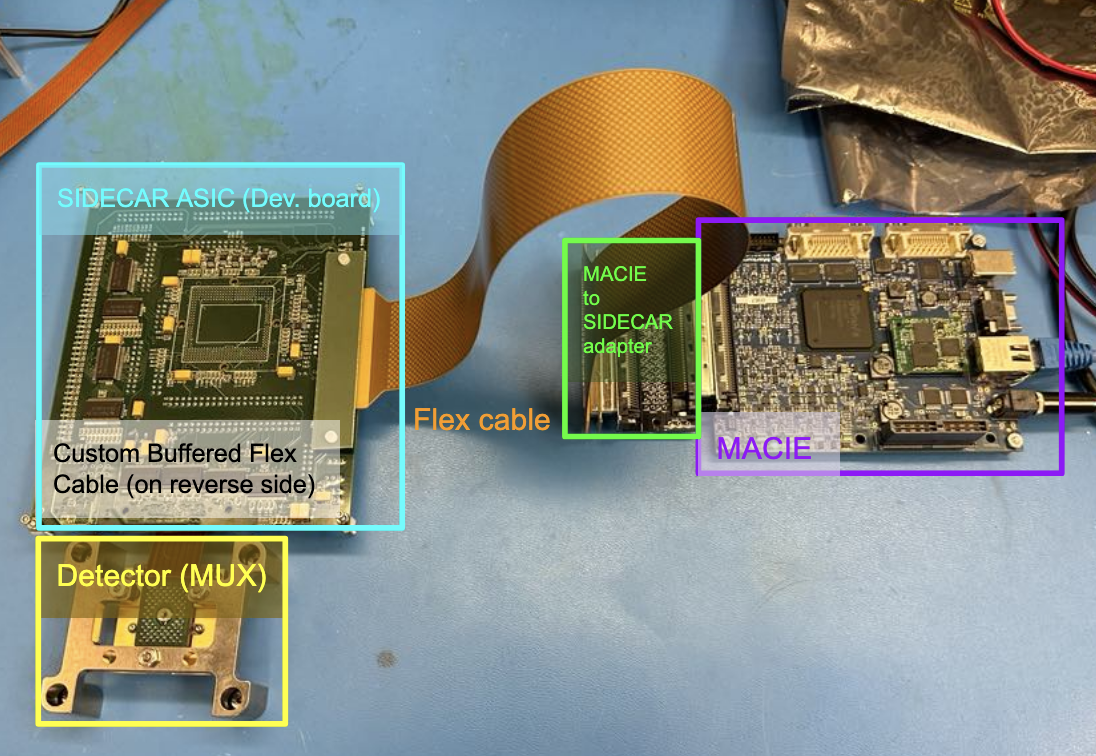}
    \caption{Room-temperature testing setup used to develop software and test individual electronic components.}
    \label{fig:mux}
\end{figure}

Before the H2RG detector enclosure hardware was ready for assembly and testing, we used a MUX to test and develop the software described in Section \ref{sec:software}. This testing setup is shown in Figure \ref{fig:mux}. Using a SIDECAR ASIC development board, additional flexible cables from previous IR Lab/UCO projects, and the IR lab MUX, we were able to test the buffer board provided by UCO while developing the SCALES software. 
The basic readout of both detectors was validated via warm testing in the UCLA IR lab before beginning any tests at the detector's true operating temperature. 

\subsection{Thermal Control Validation} \label{ss:thermal}
Each detector head is controlled separately by a Lakeshore 336 temperature controller. Each controller is connected to two independent heater and temperature sensor pairs that are attached to the metal frame connected to the detector head. Copper heat sinks are present between the heaters and detectors' metal frames to ensure there is sufficient thermal contact for the heaters' ceramic surface to transfer an adequate amount of heat to the detectors. Before beginning tests with the H2RG detectors, we first performed test cooldowns in the UCLA test cryostat of the detector head external hardware with a piece of aluminum in place of the detector, to validate our thermal control. We demonstrated our ability to satisfactorily and stably cool the detector head and then proceeded to  cryogenic testing of the H2RG detectors. Using the heater and temperature sensors, we are able to stably cool and warm the detector head at a rate of 0.5 K per minute. During tests in the UCLA cryostat, we hold the detector itself at a steady temperature, typically 35 K (although operational performance was equivalent at temperatures between 35 and 39 K). We also track the temperature of the exterior of the detector head near the ASIC, which typically reached a maximum temperature of 110K during continuous operation. This demonstrates the importance of the thermal break provided by the insulating layer of G10 between the portions of the detector head that holds each of these electronic components.

\subsection{Cryogenic Testing} \label{ss:cryo}
Multiple cooldowns of each detector head were performed in the UCLA test cryostat, cooled by a closed-cycle refrigerator system. The interior of the cryostat reached a typical pressure of approximately 10 microTorr in each test, with a bench temperature of approximately 80 K. The detector was held at a constant temperature during testing, typically 35 K. 

\subsubsection{Imager Testing}
Cryogenic testing of the Imager detector head was performed at UCLA during fall 2024. We began with no light source inside the cryostat and an opaque cover over the detector face, which allowed us to diagnose and remedy some potential sources of light leakage in our detector head design. On a subsequent cooldown, the opaque detector cover was replaced with a cover with a slot for a 30 $\mu$m pinhole, and an infrared source photodiode located outside that pinhole. The brightness of this source is controllable via control of the current flow to the diode, and tests were performed with different current values that correspond to different levels of illumination. We obtained data with the MSAC GUI as well as with the Python scripts using the packages described in Section \ref{sec:software}. 

Data were taken in single reset-read mode, in CDS mode, and in Up-the-Ramp mode, with ramps of varying lengths from 10-1000 seconds. These hybrid fast/slow-mode tests were repeated at master clock rates from 1.0 MHz (corresponding to a pixel clock rate of 0.2 MHz and a minimum frame time of approximately 5.25 seconds) up to 20 MHz (corresponding to a pixel clock rate of 4 MHz and a minimum frame time of approximately 0.26 seconds). We determined that the slow mode preamplifiers on the JWST-style H2RG had deprecated performance at master clock rates greater than or equal to 10 MHz, which was evident from the presence of additional structures within the frame that could not be eliminated through tuning. We then continued our tests with a focus on master clock rates from 1-9 MHz. In addition, we used the MSAC GUI to obtain slow-mode data using the 16bit slow-mode ADCs that are also present on the SIDECAR ASIC, testing pixel clock rates of 0.1 and 0.2 MHz.

\subsubsection{IFS Testing}
Cryogenic testing of the IFS detector head was performed at UCLA during the first few months of 2025. Similar to our test procedure for the Imager detector head, we first tested the detector in a dark configuration, and then in a subsequent cooldown added a 30 $\mu$m pinhole and an infrared source of variable brightness. 

Based on the findings of our clock rate investigation with the Imager detector, we focused on characterizing the fast-mode performance at master clock rates of 1.0, 2.5, 4.0, 5.0, 6.0, 7.0, 7.5, 8.0, and 9.0 MHz. We tested these clock rates in both the GUI and the custom Python software. In addition, we used the MSAC GUI to obtain slow-mode data using the 16bit slow-mode ADCs that are also present on the SIDECAR ASIC, testing pixel clock rates of 0.1 and 0.2 MHz. Data were taken in single reset-read mode, in CDS mode, and in Up-the-Ramp mode, with ramps of varying lengths from 10-1000 seconds.

\section{ANALYSIS AND RESULTS}
\label{sec:results}

\subsection{Interpolated Nonlinearity Correction}
\begin{figure}
    \centering
    \includegraphics[width=\linewidth]{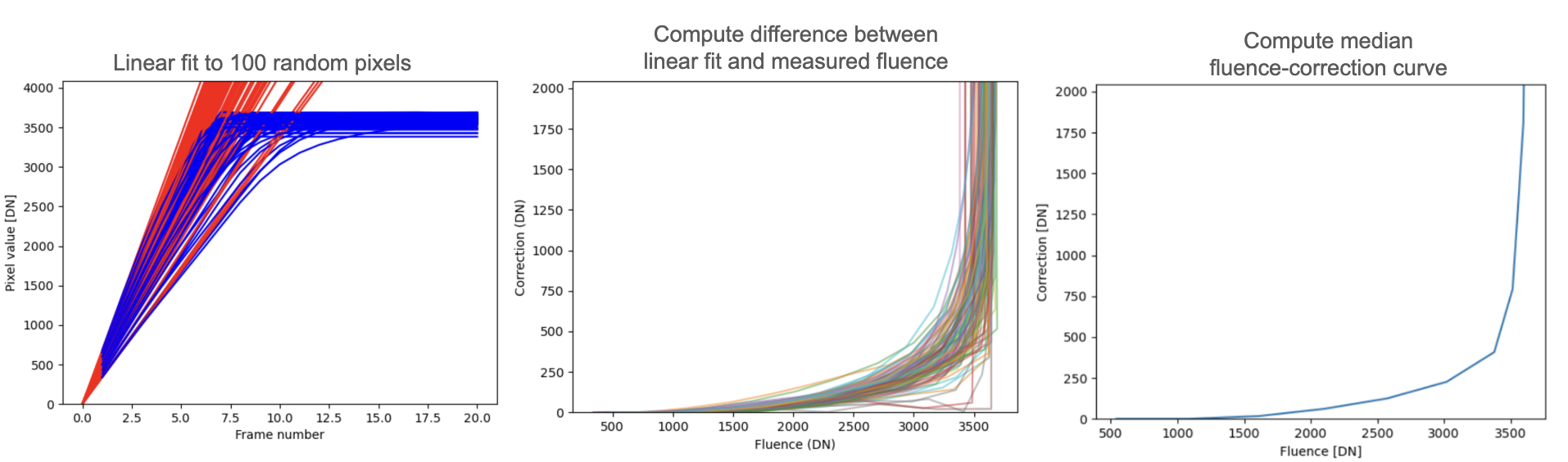}
    \caption{Basic nonlinearity correction, applied as an additive correction to each pixel based on interpolation of the median of 100 pixels' deviation from their linear fits.}
    \label{fig:nonlin}
\end{figure}

When inspecting the accumulated counts in individual pixels during illuminated ramp acquisitions, it became clear that there is significant nonlinearity, which was evident in the variation of the saturation point of each pixel. There is inherent nonlinearity in HxRG-style source follower circuits as the detector approaches saturation. It is reasonable to estimate a pixel's full well depth to be the point where the pixel response deviates 5\% from linear \cite{Blank_2011}. For a preliminary attempt at correcting this nonlinearity, we create a master correction curve that has an additive value for each measured pixel fluence (in DN), and interpolate that curve to determine the appropriate additive correction value. To do this, we perform a linear fit to a 100 pixels within a defined number of reads, and then calculate the residual between that linear fit and the measured fluence, and define that residual as the correction. Then, we take the median across the 100 pixels sampled and compute the master fluence-correction curve. This is shown in Figure \ref{fig:nonlin}.
However, this approach could obscure the pixel-to-pixel variations in nonlinearity that could be especially impactful on spectroscopic observations; for this reason, we are currently working alongside the SCALES Data Reduction Pipeline team (Unni et al., paper 13627-68, this conference) to implement a more sophisticated polynomial fit to model the nonlinearity of each pixel individually.

\subsection{Photon Transfer Curve and Gain Fitting}
\begin{figure}
    \centering
    \includegraphics[width=\linewidth]{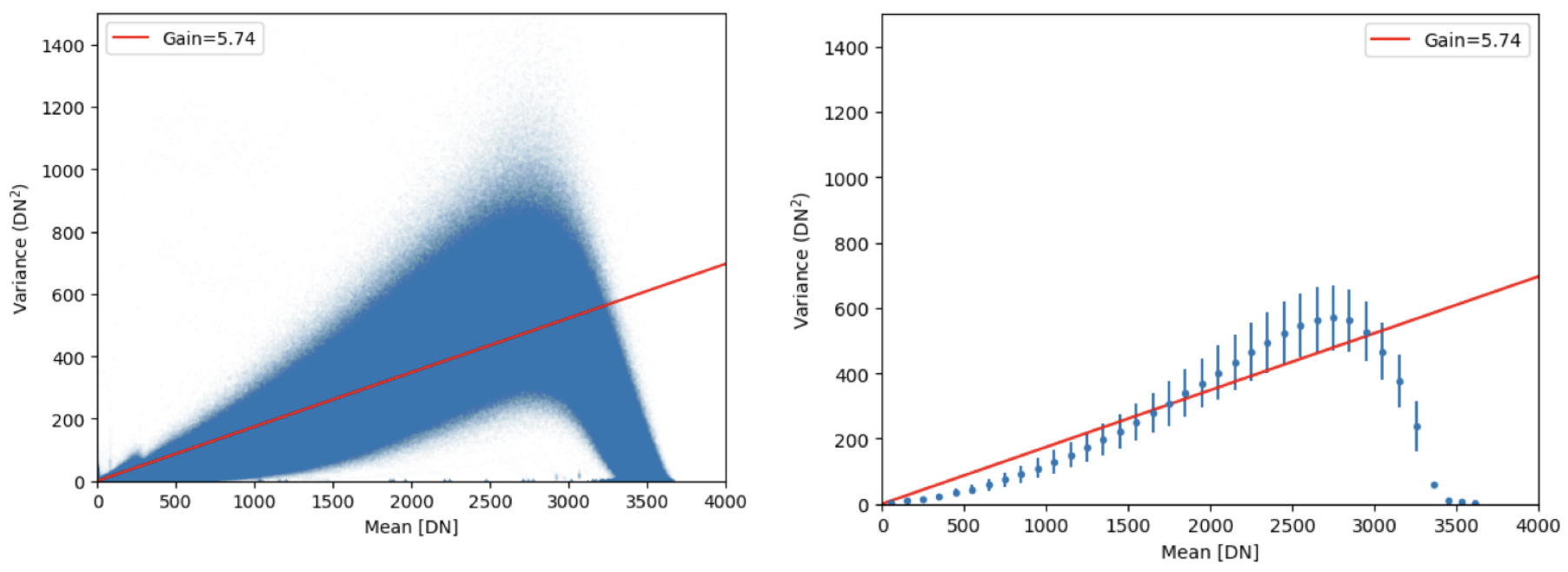}
    \caption{Photon transfer curve for the Imager detector, showing the pixelwise mean and variance measured across 5 identical ramps. A linear fit yields a slope that corresponds to a gain of 5.74 $e^-/DN$, close to the nominal value of 5.66. The panel at left shows every data point as a scatter plot, and the panel at right shows the same data binned for clarity and presented with the median absolute deviation of the variances within that bin as the width of the error bar.}
    \label{fig:PTC}
\end{figure}

Although we have tested multiple values of the preamplifier gain of the SIDECAR ASIC, we performed the majority of the tests with a specific gain setting, defined in the SIDECAR ASIC manual to correspond to a nominal gain of 5.66 $e^-/DN$. To experimentally measure the gain of the Imager detector, we create a photon transfer curve (shown in Figure \ref{fig:PTC}).  To do this, we acquire 5 identical ramps of illuminated data, cropped to only look at the illuminated portion of the field. We then apply the nonlinearity correction described in the previous subsection. For each read in the length of the ramp, we compute the mean and variance of each illuminated pixel across the 5 distinct ramps. Plotting this mean and variance yields the scatter plot in the left panel of Figure \ref{fig:PTC}. Similar analysis is ongoing for the IFS detector.  

\subsection{Reference Pixel Subtraction}
\begin{figure}
    \centering
    \includegraphics[width=\linewidth]{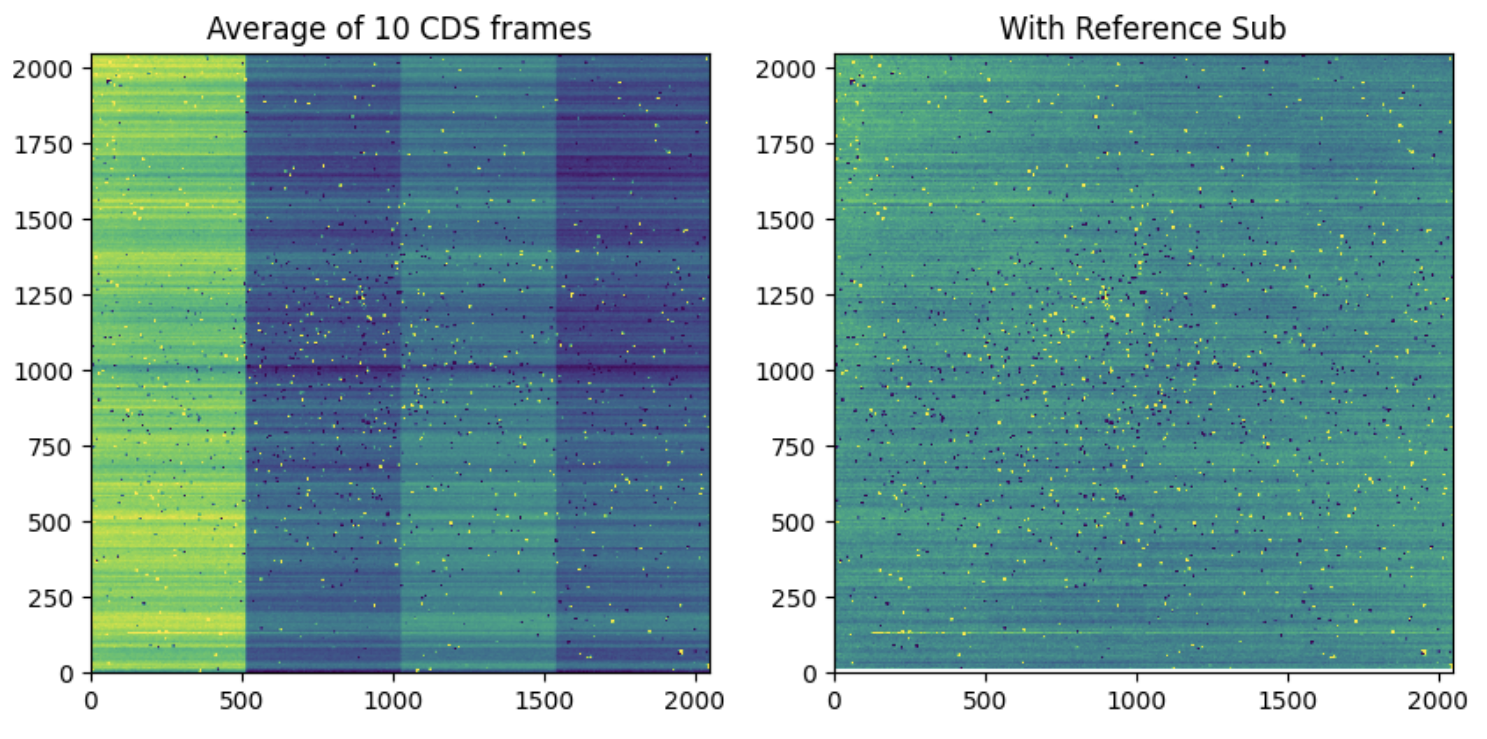}
    \caption{6 MHz Imager CDS results from the UCLA test cryostat, presented without (left) and with (right) the reference pixel subtraction implemented via Python. Data are presented with the same color scale.}
    \label{fig:RefPix}
\end{figure}

The H2RG detectors have 4 reference pixels bordering each edge of the frame. These non-photosensitive pixels are used to perform reference pixel subtraction. In the MSAC GUI, this is implemented automatically, but implementation in the Python code is ongoing. Figure \ref{fig:RefPix} shows an example of averaged CDS results from the Imager detector, obtained via Python, shown with and without the reference pixel subtraction applied. This demonstrates both the output-level correction that balances the levels in the four output channels, as well as the row-level correction that reduces the horizontal striped structure. 

\subsection{CDS Noise Characterization} \label{ss:cdsNoise}

To estimate the noise performance of each detector, we compute the standard deviation of every pixel in the frame across an identical set of $N_{frames}$ CDS result frames from the MSAC GUI. We then take the median of all the pixelwise standard deviations and present it in Table \ref{tab:cdsnoise}.
\begin{table}[ht]
\caption{CDS Noise estimates as calculated in Section \ref{ss:cdsNoise}. For all values marked with $^*$, the number of frames ($N_{frames}$) used is 30; all other values are computed with $N_{frames} = 10$. Values for both detectors are converted from DN to electrons using the conversion gain found by fitting the Imager photon transfer curve shown in Figure \ref{fig:PTC}.} 
\label{tab:cdsnoise}
\begin{center}       
\begin{tabular}{|l|l|l|l|l|l|} 
\hline
\rule[-1ex]{0pt}{3ex}  Clock Rate & Clock Rate & Imager CDS & Imager CDS & IFS CDS & IFS CDS \\
\rule[-1ex]{0pt}{3ex} (Master, MHz) & (Pixel, MHz) & Noise est. (DN) &  Noise est. ($e^-$) & Noise est. (DN) & Noise est.  ($e^-$)\\
\hline
\rule[-1ex]{0pt}{3ex} 5.0 & 1.0 & 1.751$^*$ & 10.01$^*$ & 0.598 & 3.432  \\
\hline
\rule[-1ex]{0pt}{3ex}  7.5 & 1.5 & 0.857 & 4.919 & 0.675 & 3.875 \\
\hline
\rule[-1ex]{0pt}{3ex} 8.0 & 1.6 & 0.775 & 4.449 & 0.676 & 3.880 \\
\hline
\rule[-1ex]{0pt}{3ex}  9.0 & 1.8 & 1.394 & 8.002 & 0.656 & 3.765 \\
\hline

\end{tabular}
\end{center}

\end{table}

\subsection{CDS Results}

\begin{figure}
    \centering
    \includegraphics[width=\linewidth]{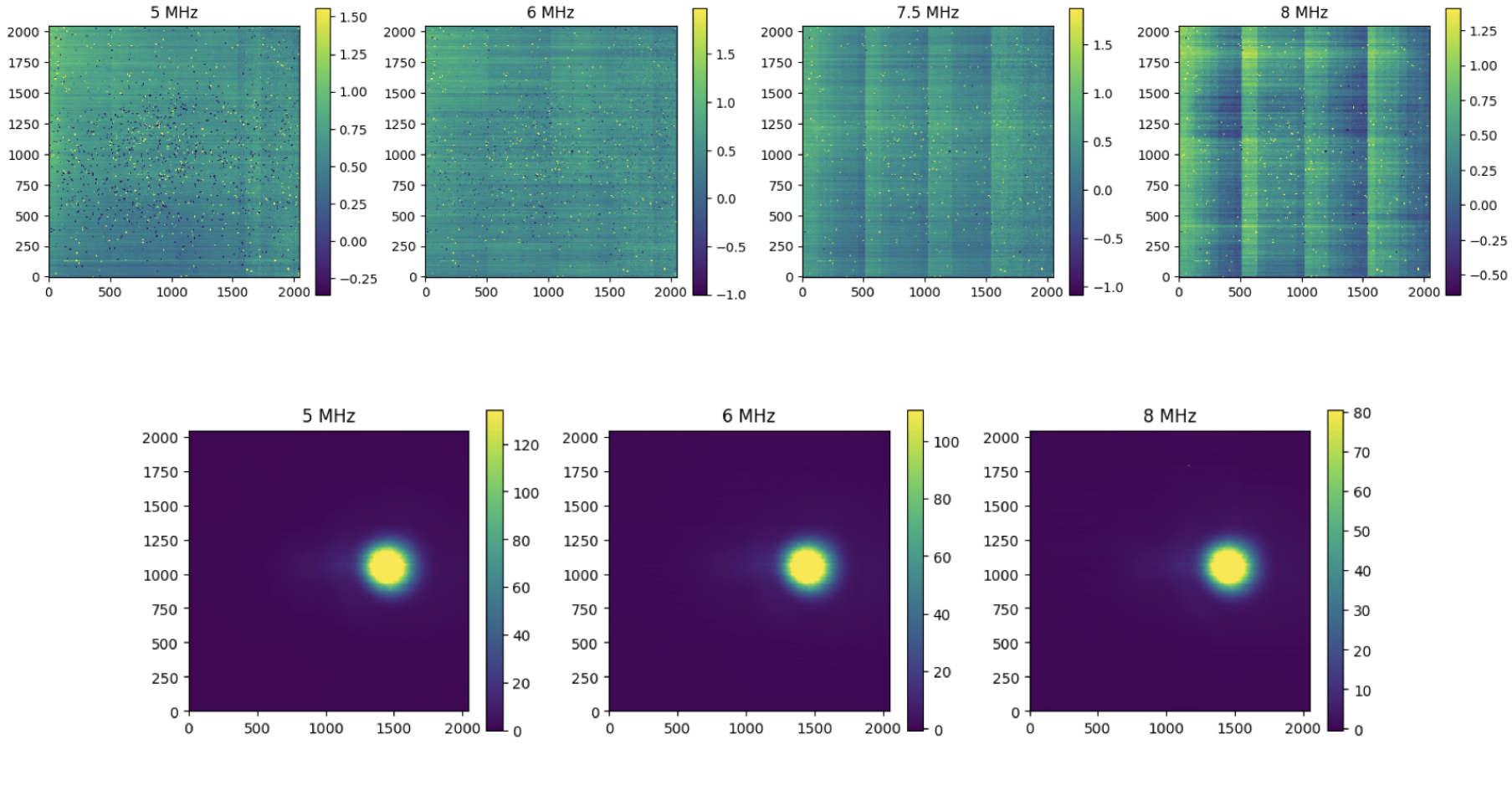}
    \caption{Dark (top row) and illuminated with 30 micron pinhole (bottom row) data from the Imager detector, captured in the UCLA test cryostat via Python on November 14, 2024. Data are shown at a variety of master clock rate values from 5.0 MHz (corresponding to a minimum frame time of 1.05 seconds) to 8.0 MHz (corresponding to a minimum frame time of 0.66 seconds.) These data are shown with reference pixel subtraction applied. All illuminated data were taken with the same source brightness, and all data are plotted with the colorbar's limits set to their 1st and 99th percentile.}
    \label{fig:ImData}
\end{figure}

\begin{figure}
    \centering
    \includegraphics[width=\linewidth]{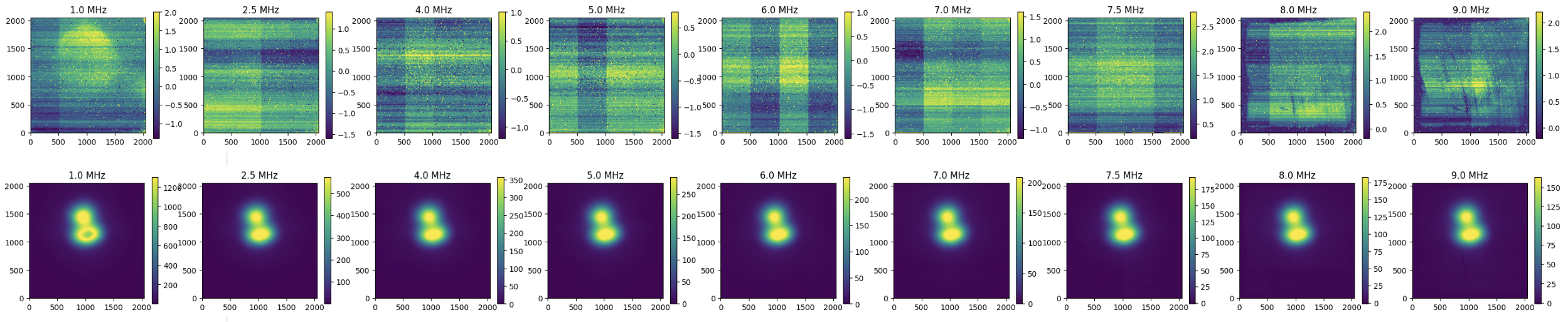}
    \caption{Dark (top row) and illuminated with 30 micron pinhole (bottom row) data from the IFS detector, captured in the UCLA test cryostat via Python on June 9, 2025. Data are shown at a variety of master clock rate values from 1.0 MHz (corresponding to a minimum frame time of 5.24 seconds) to 9.0 MHz (corresponding to a minimum frame time of 0.58 seconds.) These data are shown with reference pixel subtraction applied. All illuminated data were taken with the same source brightness, and all data are plotted with the colorbar's limits set to their 1st and 99th percentile.}
    \label{fig:IFSdata}
\end{figure}

To demonstrate the visual performance of each detector, we present dark and illuminated CDS frames obtained via Python for both the Imager (see Figure \ref{fig:ImData}) and IFS (see Figure \ref{fig:IFSdata}) detectors. In Figure \ref{fig:ImData}, inspecting the top row of dark CDS results, it is clear the reference pixel subtraction adequately removes horizontal and vertical structure at master clock rates of 5 and 6 MHz. However at 7.5 and 8 MHz, in our Imager tests at UCLA, we were unable to optimize detector parameters to eliminate the intrareadout structure that seems to result from the fast clocking of the slow mode preamplifiers. (However, in subsequent tests we have found evidence that we have been successful in reducing this effect both with the IFS detector when tested at UCLA and the Imager detector when tested at UCSC.) The bottom row of Figure \ref{fig:ImData} shows the illuminated CDS results at a master clock rate of 5, 6, and 8 MHz, with the same source brightness. 


In Figure \ref{fig:IFSdata}, dark (top row) and illuminated (bottom row) CDS results are shown for the complete range of master clock rates tested in this work (1.0 - 9.0 MHz). Improvements to detector and ASIC tuning were made in the approximately six months between Imager and IFS characterization, which resulted in a change in the forms of substructure that are evident in the dark data. Unlike the noisy intrareadout patterns caused by fast clocking shown in Figure \ref{fig:ImData}, the patterns shown in the dark IFS data (Figure \ref{fig:IFSdata}) are likely to result from either imperfect reference pixel subtraction, or from additional sources of noise (e.g. 1/f noise, or other sources of correlated noise \cite{rauscher24}) for which we are not yet correcting. Additionally, the dark data shown here were obtained with the 30 $\mu$m pinhole in place and the infrared source powered off. At the slowest clock rate (master clock rate of 1 MHz, corresponding to a single frame time of 5.24 seconds), enough stray light was able to enter the pinhole to create a noticeable circular glow (see the upper left frame in Figure \ref{fig:IFSdata}.) Our placement of the light source relative to the pinhole was imperfect in this test, causing the double-lobed appearance of the pinhole's illumination. In the slowest tested clock rate (1 MHz), the center of the brighter lobe saturated during the CDS acquisition, resulting in the dark oversubtracted region in the lower lobe.  All the illuminated data shown for the IFS detector were acquired at the same source brightness.

\section{DISCUSSION}
\label{sec:disc}

Our tests have validated the basic performance of the detectors, with a clear response to photon signals and a noise profile generally consistent with that found in the NASA Goddard test data characterizing these detectors \cite{rauscher14_newHxRG}.
In coming tests, one of our crucial areas of focus will be quantifying the persistence of the detectors. Specifically, the H2RG detector used as the Imager in SCALES was rejected from the JWST mission due to not meeting their stringent persistence requirements for space operation. Tests at UCLA did not show significant evidence of persistence, but we will be conducting short-, medium-, and long-scale tests during the upcoming tests at UCSC to watch for persistent signals from bright sources. 

Analysis of the existing data is ongoing. We aim to produce more photon transfer curves, like the example for the Imager detector at a  master clock rate of 5 MHz shown in Figure \ref{fig:PTC}. This will allow us to validate the conversion gain for both detectors and in a variety of operational modes. Additionally, we would like to more robustly analyze the correlation of a pixel with its neighbors, as well as correlations between pixels in different output channels. This is crucial for determining the effect of crosstalk at fast clock rates, and therefore may provide a limit on our fastest effective clock rate. Once our characterization of these essential detector properties (e.g. noise profile, conversion gain) is complete, we plan to integrate our results into both the \textit{scalessim} Python package used for SCALES observation planning \cite{ScalesSim_2020}, and the SCALES data reduction pipeline.

In the UCLA tests of both H2RG detectors for SCALES, we successfully demonstrated clocking at pixel clock rates up to 1.8 MHz, corresponding to a minimum frame time of 0.58 seconds. This crucially increases the magnitude range of the brightest targets accessible to SCALES, as well as minimizing the impact of the infrared sky brightness. SCALES is presently undergoing integration and testing in advance of a delivery to Keck Observatory for commissioning in early 2026.  SCALES' ambitious science campaign will be made possible by the sensitive, low noise performance of our hybrid fast-slow readout scheme for these unique H2RG detectors.



\acknowledgments 
The firmware upgrade described in this proceedings was funded by the Mt. Cuba Astronomical Foundation.  Major grants to the SCALES project include NSF Grant 2216481 as well as grants from the Heising-Simons Foundation, the Mt. Cuba Astronomical Foundation, and the Alfred P. Sloan Foundation.  We are also grateful to the Robinson family and other private supporters, without whom this work would not be possible. P.B. is a member of UAW 4811, and her work on SCALES is funded by University of California Observatories.

\bibliography{biblio} 
\bibliographystyle{spiebib} 

\end{document}